# Gender Imbalance and Spatiotemporal Patterns of Contributions to Citizen Science Projects: the case of Zooniverse

**Khairunnisa Ibrahim[1], Samuel Khodursky[2,3], Taha Yasseri[1,4,5,6*]**

[1] Oxford Internet Institute, University of Oxford, Oxford, UK
[2] Laboratory of Evolutionary Genetics and Genomics, The Rockefeller University, New York, USA
[3] Department of Physics, University of Oxford, Oxford, UK
[4] School of Sociology, University College Dublin, Dublin, Ireland
[5] Geary Institute for Public Policy, University College Dublin, Dublin, Ireland
[6] Alan Turing Institute for Data Science and AI, London, UK

*Corresponding Author: taha.yasseri@ucd.ie*

## Abstract

Citizen Science is research undertaken by professional scientists and members of the public collaboratively. Despite numerous benefits of citizen science for both the advancement of science and the community of the citizen scientists, there is still no comprehensive knowledge of patterns of contributions, and the demography of contributors to citizen science projects. In this paper we provide a first overview of spatiotemporal and gender distribution of citizen science workforce by analyzing 54 million classifications contributed by more than 340 thousand citizen science volunteers from 198 countries to one of the largest citizen science platforms, Zooniverse. First we report on the uneven geographical distribution of the citizen scientist and model the variations among countries based on the socio-economic conditions as well as the level of research investment in each country. Analyzing the temporal features of contributions, we report on high "burstiness" of participation instances as well as the leisurely nature of participation suggested by the time of the day that the citizen scientists were the most active. Finally, we discuss the gender imbalance among citizen scientists (about 30% female) and compare it with other collaborative projects as well as the gender distribution in more formal scientific activities. Citizen science projects need further attention from outside of the academic community, and our findings can help attract the attention of public and private stakeholders, as well as to inform the design of the platforms and science policy making processes.

## Introduction

The growth of online collaboration between scientists and the public on research projects in recent years have led to numerous discoveries at unprecedented rates in various fields of science. These projects called *citizen science*, are part of scientific initiatives that involve non-scientists in research work. Not only does citizen science facilitate new breakthroughs [1], they do so thanks to voluntary contributions. The collective efforts of citizen scientists, as these volunteers are known, to one of the citizen science platforms

Zooniverse[1], was estimated to be worth US$1.5 million in 2015 by Sauermann & Franzoni [2] and represent the equivalent of 34 years' full time service [3] and 50 years of non-stop research [4].

It is not only the scientific community that benefits from citizen science; Research has shown that volunteers gain learning opportunities, positive attitudes towards science and the ability to participate in research [5, 6, 7, 8] as well opportunity for socializing and participating in an online community [6, 9]. In addition to making science more open and accessible, online citizen science accelerates research by leveraging human and computing resources [10, 11, 12], tapping into rare and diverse pools of expertise [2, 13], providing informal scientific education and training, motivating individuals to learn more about science [14]. Given these advantages, there has been a surge in research that seek to understand the activities and drives of both the scientists and the citizen scientists involved. As Rotman et al. note, by "understanding the shared and unique motivations of these two groups" citizen science developers can design "the technical and social infrastructures needed to promote effective partnerships" [15].

This paper contributes towards those efforts by focusing on the geographical and temporal patterns of participation in the multi-project citizen science platform, the Zooniverse. Previous studies have focused on motivations of citizen scientists [11, 12], the frequency, productivity and intensity of volunteers' engagement [18] and preference between contributing a classification – defined as a unit of task completed in a citizen science project [4] – and participating in a discussion forum [11, 7]. Others have looked at factors that promote or hinder participation, such as whether taking part in discussion forums have any influence on classification activities [19] and why volunteers only 'drabble' or contribute occasionally, or even drop-out [6]. To date, however, there is little research on where volunteers come from, and when they are active, although Ponciano et al. [18] and Sauermann and Franzoni [2] do briefly examine temporal patterns with regard to frequency of activity over certain time periods.

Access to the technologies and skills to participate in online activities and for data creation and sharing is becoming ever more ubiquitous. Consequently, significant amount of research has been undertaken to discover the spatiotemporal patterns of these activities towards better understanding of human behavior, society and technology. A common finding among geographical studies of online activity is the disproportionate spatial distribution of both participation and data creation. These "stark core-periphery patterns" [20] reaffirm the dominance of resources and power in *the West*, and the lack of both in *the global South*. These are patterns that have been mapped and replicated across various online platforms, such as the photo-sharing site Flickr [21], the online encyclopedia Wikipedia [20], Twitter [22] and Google Maps [23]. Studies that examine where and what data are and are not available online thus reveal social realities, processes and divisions [19, 20, 21].

---

[1] https://www.zooniverse.org/

Temporal patterns of online activity have also been extensively studied in the field of *Human Dynamics*. Recent research include the editing behavior of editors on OpenStreetMap [27], Wikipedia [28], check-ins on the location-based social network Foursquare [29], phone calls [30], text messages [31], and traffic on the video-sharing platform Youtube [32]. These studies helped discern human tendencies at different time-scales. For instance, Noulas et al., by comparing Foursquare activity during the weekday and weekends [29], found distinct variance in when and where people check in. Similarly, Kaltenbrunner et al. discovered that people were considerably more active on the technology-news website Slashdot during the weekday compared to weekends [33]. Temporal patterns were also used to infer other information not readily available, such as the geographical location of editors of different language editions of Wikipedia [28].

According to a survey by the Zooniverse, about 30% of their respondents are from the UK, 35% from the US and the rest from other parts of the world [34]. Our paper investigates the patterns of activity in the Zooniverse between 2009 and 2013 by analyzing 54 million classifications made in 17 different citizen science projects by more than 340,000 volunteer citizen scientists. Our focus is on the spatial and temporal patterns of activity; we are interested in where volunteers contribute to citizen science from, and during what times of the day and week they are most active. We seek to answer two basic questions:

    1) Where do volunteers contribute from, and when?
    2) What are some of the reasons for these patterns?
    3) What is the gender distribution among the contributors?

## Materials and Methods

Data overview

This study utilizes a dataset of 54 million classifications made in 17 different citizen science projects in the Zooniverse made between November 2009 and June 2013 (see Supplementary Information for a list and timeline of the projects). The classifications were produced by more than 340,000 volunteers from 198 different countries. Each classification record includes a unique classification id, the volunteer's distinct user-id and their approximate geographical location, the timestamp of the classification and the project to which the classification is made. The records also include gender information. However, volunteers are not required to identify as either male or female, so gender information in the dataset is derived from a separate analysis of the first names that they used to register using the Gender API.[2] The complete dataset is available at https://doi.org/10.5281/zenodo.583182.

## Geographical analysis

Country data on citizen science is obtained by extracting the numbers of unique volunteers and the aggregate number of classifications made per country. The dataset uses two-letter codes to identify countries (e.g. GB for United Kingdom, US for United States). Country-level socioeconomic data are obtained from the World Bank. Four types of socioeconomic variables are taken from the World Bank database: population, income (GDP per capita), Internet connectivity (number of Internet users) and education (primary and secondary school enrolment). Table 1 provides an overview of the socioeconomic variables.

**Table 1**: Socioeconomic variables Adapted from the World Bank[3]

| **Variable** | **Description** |
|---|---|
| Population | Includes all residents "regardless of legal status or citizenship" but excludes refugees who are not "permanently settled in the country of asylum". The latter are generally considered part of the population of their country of origin. The values are mid-year estimates of the population. |
| GDP per capita | The Gross Domestic Product divided by the midyear population. GDP is the "sum of gross value added by all resident producers in the economy plus any product taxes and minus any subsidies not included in the value of the products". Data are in expressed current US dollars. |
| Internet users per 100 people | "People with access to the worldwide network" |
| Gross primary school enrolment (%) | Total is the total enrollment in primary education, regardless of age, expressed as a percentage of the population of official primary education age. GER can exceed 100% due to the inclusion of over-aged and under-aged students because of early or late school entrance and grade repetition. |
| Gross secondary school enrolment (%) | The total is the total enrollment in secondary education (all programmes), regardless of age, expressed as a percentage of the population of official secondary education age. GER can exceed 100% due to the inclusion of over-aged and under-aged students because of early or late school entrance and grade repetition. |

---

[2] https://gender-api.com/
[3] http://data.worldbank.org

The three variables for scientific culture are also obtained from the World Bank. An overview is provided in Table 2.

**Table 2:** Scientific culture variables Adapted from the World Bank

| Variable | Description |
|---|---|
| Research and development expenditure (% of GDP) | Figures indicate the "current and capital expenditures (both public and private) on creative work undertaken systematically to increase knowledge, including knowledge of humanity, culture, and society, and the use of knowledge for new applications" and includes "basic research, applied research, and experimental development". |
| Researchers in R&D (per million people) | Figures indicate the number of professional involved in "the conception or creation of new knowledge, products, processes, methods, or systems and in the management of the projects concerned", including postgraduate PhD students. |
| Scientific and technical journal articles | Figures indicate the number of scientific and engineering articles published in the fields of physics, biology, chemistry, mathematics, clinical medicine, biomedical research, engineering and technology, and earth and space sciences. |

Temporal analysis

The daily cycles of activity are examined for the 20 countries most active in the Zooniverse. The standardized time is taken to be the Greenwich Meridian Time (GMT). All other countries' data were adjusted to the GMT. Several countries, most notably the US, Canada, Australia, the Russian Federation and Brazil, span several time zones. The decision was taken to adjust the data using an average of +3 and -3 hours respectively for the Russian Federation and Brazil. For the US, Canada and Australia, the data for various regions are aggregated and adjusted according to their various time zones. The hourly contributions by region are then added to represent the countries' overall activity per hour.

The circadian rhythms for each country ($Circadian_{\Delta t}$) are obtained by dividing the number of classifications made at any particular hour ($Classifications_{\Delta t}$) with the total number of classifications made within that country ($Classifications$).

$$Circadian_{\Delta t} = \frac{Classifications_{\Delta t}}{Classifications}$$

Meanwhile, the universal curve is the ratio of the total number of classifications per hour made by the 20 countries over the overall number of classifications from the 20 countries made throughout the 24-hour period.

$$Universal = \frac{\sum_{20} Classifications_{\Delta t}}{\sum_{20} Classifications}$$

## Results

Spatial patterns of citizen science activity

The Zooniverse volunteers are distributed worldwide, but the rate and preferences of their engagement vary. The majority of volunteers and activity come from North America and Western Europe. Even when normalized for population, the same pattern holds (Figure 1), indicating that the majority of citizen scientists are to be found in Western, developed countries, regardless of the numbers of people within each country. These spatial patterns of engagement and activity reflect other geographical trends in online activity. Similar patterns are observed on Flickr [21], the online encyclopedia Wikipedia [20], Twitter [22] and Google Maps [23]. As with these other online platforms for creating and sharing knowledge, The Zooniverse receives intense bursts of activity in the global North, while other parts of the world, particularly the African continent, remain seemingly disconnected from these platforms.

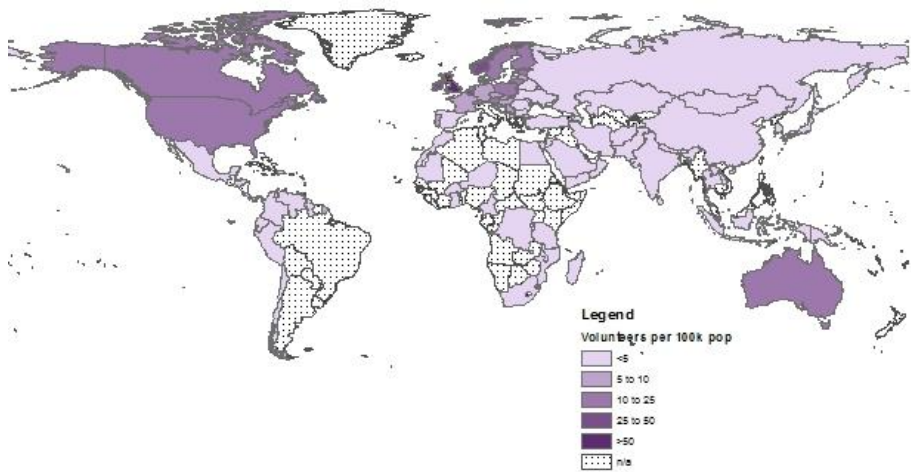

**Figure 1:** Spatial distribution of citizen scientists normalized by population

What are the reasons for the uneven geographies of engagement in citizen science? We considered various factors, which fall into two categories: ***socio-economic*** and ***scientific culture***.

*Socioeconomic* factors can indicate the propensity of the average individual in a country to engage in citizen science. Citizen science requires the voluntary participation of many people, which is more readily provided if the volunteers are relatively well-off, have easy access to the Internet, and have ample free time. Our analysis of the temporal patterns of activity, shared later in the paper, also point to citizen science as being primarily a leisurely pursuit, taken up after the normal workday is over. Finally, the nature of the citizen science projects indicates at least a certain level of education on the part of the volunteer, in order for them to understand and be interested in the projects.

A country's *scientific culture* can also be a good indicator for its residents' involvement in citizen science. Countries that spend more on research and development, cultivates more researchers, and produce more scientific outputs signal a vibrant scientific community that encourages experimentation and learning. This positive inclination towards knowledge production may also extend to those outside the formal institutions of science, for even when individuals move into non-scientific careers, their early education will most likely have included science subjects.

Using the latest available indicators from the World Bank, we analyzed the relationship between engagement in citizen science, defined as the number of contributions made from each country. The socioeconomic variables include population, GDP per capita, Internet connectivity (the number of Internet users per 1000 people) as well as primary school enrolment. Countries' scientific cultures are measured by the expenditure on research and development initiatives, the number of researchers employed in various sectors and the overall number of publications in science and technology journals and books.

The socioeconomic analyses confirm a positive link between citizen science and socioeconomic indicators; that is, the wealthier and connected a country is and the more educated its residents, the higher the rate of participation and contribution. Table 3 shows the Pearson correlation results of the socioeconomic analysis (controlling for population). On the whole, most of the variables have a positive relationship with citizen science engagement, although the magnitude of these relationship varies.

**Table 3:** Bivariate correlations (Pearson correlation) among citizen science and socioeconomic variables

|  | **Internet Users** | **GDP per capita** | **Primary education** |
|---|---|---|---|
| **volunteers (log)** | 0.72 | 0.58 | 0.51 |
| **contributions (log)** | 0.80 | 0.66 | 0.41 |

Having these primary results in hand, we conducted tests on a number of regression models on socioeconomic indicators. Our analysis of just over 200,000 cases, each representing a volunteer from one of 90 countries (Table 4), revealed that over 85% of the variation in classifications made per country can be explained by the combination of all five factors: population, GDP per capita, Internet users per 100 people, and gross school enrolment (Model 1). When the educational variables are excluded (Model 2), the adjusted $R^2$ is still fairly high at 0.80. However, when only Internet connectivity and school enrolment are considered (Model 3), the outcome is still considerable at 0.66. These results indicate that taken separately, each factor has varying rates of influence on citizen science, but in tandem they explain the major bulk of engagement in the Zooniverse.

**Table 4** Models of socioeconomic factors for the geographies of citizen science

|  | Model 1 | | Model 2 | | Model 3 | |
| --- | --- | --- | --- | --- | --- | --- |
|  | **B** | **SE** | **B** | **SE** | **B** | **SE** |
| **Constant** | -12.96*** | 0.028 | -11.60*** | 0.021 | | |
| **Population (Log)** | 0.856*** | 0.002 | 0.935*** | 0.002 | | |
| **GDP per capita (Log)** | 0.328*** | 0.007 | -0.465*** | 0.007 | | |
| **Internet connectivity (Log)** | 4.944*** | 0.018 | 6.786*** | 0.019 | 7.010*** | 0.012 |
| **SSE (Log)** | -0.014*** | 0.000 | | | -0.037*** | 0.000 |
| **PSE (Log)** | 0.031*** | 0.000 | | | -0.008*** | 0.000 |
| **adjusted $R^2$** | 0.856 | | 0.801 | | 0.658 | |

*** $p<0.01$

Residuals from the Model 1 for each country are shown in Figure 2. Where we see under-representation from North-American countries and over-representation from some of the developing countries in Asia and Africa.

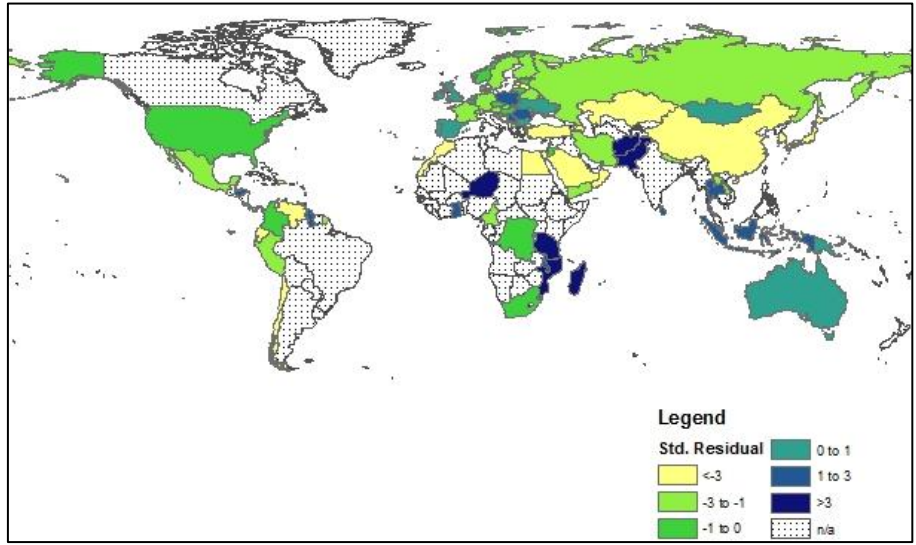

**Figure 2:** Standard residuals for Model 1 (pop, GDP per capita, Internet users and education)

Our analysis of the influence of scientific culture on citizen science engagement indicate that more scientifically active countries are more likely to be involved in citizen science. We ran a correlation test of over 21,000 cases from 56 countries with available science data - spending on research and development, the number of researchers and publications - from the World Bank. Table 5 shows the result of this analysis. All three variables show a positive correlation, with R&D budget and science publications being highly correlated at over 70% and 80% respectively.

**Table 5:** Pearson-moment correlation among citizen science variables and science indicators

|  | **R&D budget (log)** | **# of researchers (log)** | **S & T publications (Log)** |
|---|---|---|---|
| **Volunteers (log)** | 0.75 | 0.48 | 0.80 |
| **Contributions (log)** | 0.78 | 0.56 | 0.82 |

A linear combination of the three variables contribute to 73.2% of the variation in citizen science activity (Table 6). The results indicate that while research spending and the availability of human resource in science are necessary, it is the outcomes of these investment and resources that have greatest impact on involvement in citizen science. In other words, researchers and funding agencies should aim for increasing their publication record as this research shows that greater publications can influence more involvement in science even outside academia and industry.

**Table 6:** Least squares linear regression predicting log (contributions) per country

|  | B | SE | 95% CI | |
|---|---|---|---|---|
| **Intercept** | 1.545*** | 0.033 | 1.481 | 1.609 |
| **Budget (log)** | -0.298*** | 0.005 | 1.124 | 1.140 |
| **Researchers (log)** | 0.832*** | 0.004 | -0.308 | -0.288 |
| **S&T publications (log)** | 1.132*** | 0.004 | 0.824 | 0.840 |
| **Adjusted $R^2$** | 0.732 | | | |

*\*\*\* p<0.01*

In Figure 3, the residuals of the model are shown which is similar to the pattern in Figure 2, with the exception of more contributions from the USA compared to the model prediction.

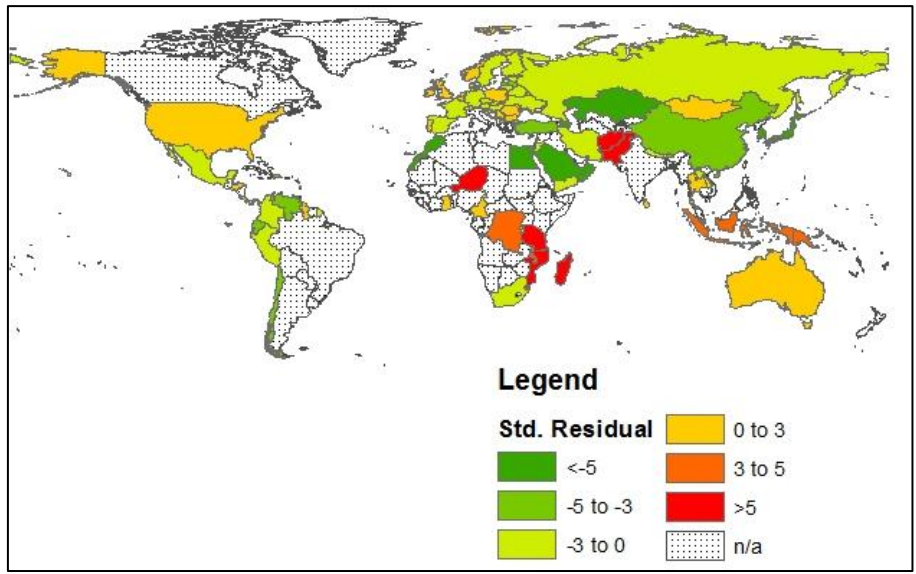

**Figure 3:** Standard residuals for scientific culture model

Temporal patterns of citizen science activity

Distribution of Time between Contributions to Zooniverse

Many social and complex systems variables exhibit fat-tailed distributions. The wealth of people [35], the populations of cities [36], and the number of citations to papers [37], are all quantities that are distributed very far from a normal distribution, with few instances of very large values an many instances of small values. The distribution of time between successive contributions by each citizen scientist to Zooniverse is shown in Figure 4. The fat-tailed distribution of the time intervals is in line with previous reports on similar patterns among Wikipedia edits [38], emails sent/received [39], phone calls made [40], and many more examples of *unscheduled* human activities. This type of distribution of time-intervals is a fingerprint of *bursty* behaviour, in which sessions of many activities are often followed by long waiting times between sessions [41].

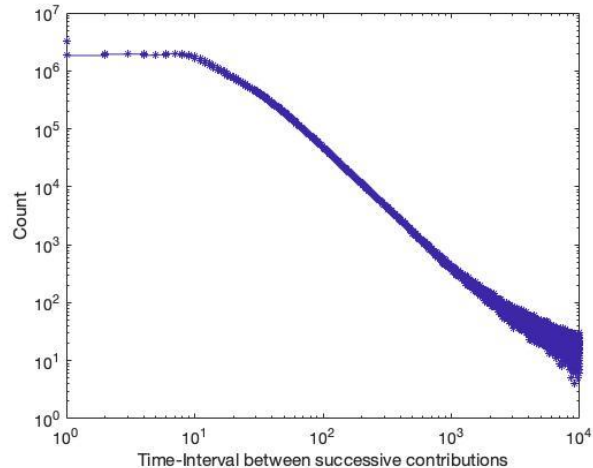

**Figure 4:** Distribution of time-intervals between successive contribution by the same citizen scientist.

Circadian patterns of activity

We extracted the number of classifications made by volunteers in one-hour windows throughout the day, normalized the values for each window to the total number of contributions throughout the 24-hour period, and plotted the circadian charts of activity. Figure 5 shows the circadian patterns of the 20 most active countries in Zooniverse.

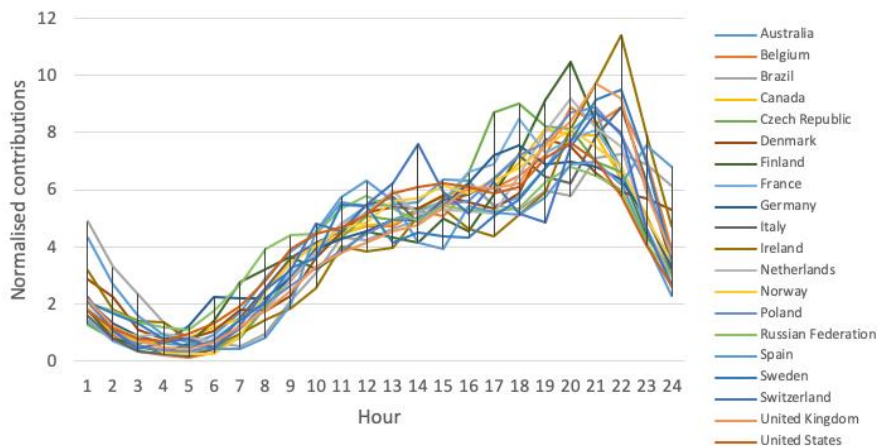

**Figure 5:** Circadian rhythms of contributions from 20 countries to the Zooniverse

We also calculated the average temporal patterns for all the selected countries to produce the universal curve shown in Figure 6. This curve represents the general pattern of activities over the 24-hour period. It shows that overall, activity in the Zooniverse starts to pick up gradually after 5am. The number of classifications continue to increase throughout the day, with small dips in activity in the afternoon (around 2pm) and early evening (around 6pm). The rate of activity peaks at 9pm, and falls in volume in the following hours.

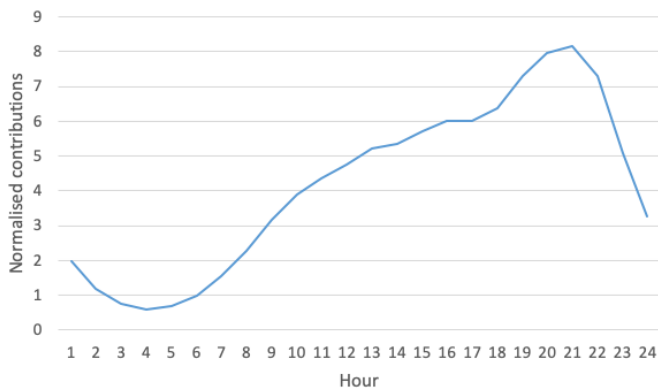

**Figure 6:** The universal curve showing the average contributions per hour for the 20 countries

This universal circadian pattern for the 20 selected countries indicates that while classifications were made at all hours of the day, most of the activity took place at night. The peak hour of 9pm is typically leisure time for most households. Assuming that most people only have about an hour or less to contribute to the Zooniverse, this pattern corresponds with findings from a survey on the Zooniverse community, in which the majority of respondents selected the option 'Only when I have spare time' in answer to the query on when they tend to classify [34]. It also reflects the tendency of people to collectively contribute to a pursuit *en masse* when they are in possession of free time and the resources to do so, a phenomenon that Shirky calls 'cognitive surplus' [42].

Time of first contributions

Two-thirds of volunteers to the Zooniverse make only one classification and do not return [18]. Given that such a large proportion of volunteers do not continue to contribute, it is important to know when most people first come to the project, so that project managers can identify measures to retain participation, and schedule them at the appropriate times. We investigated time of first contribution for the 20 countries and found that show that on average there is a marked concentration of volunteers making an initial contribution at 9pm (Figure 7). This corresponds with the universal circadian pattern above, where 9pm is shown to experience the highest activity rates in the 20 countries. Here we see a sharper peak around 9pm compared to the overall activity curve. Citizen science in general seems to be a leisure activity, undertaken when individuals have time to spare. The prevalence of nighttime activity might also be due to individuals first learning about the various projects through social media, whether directly from the Zooniverse or others [2], when they become connected to these informational outlets after work and dinner.

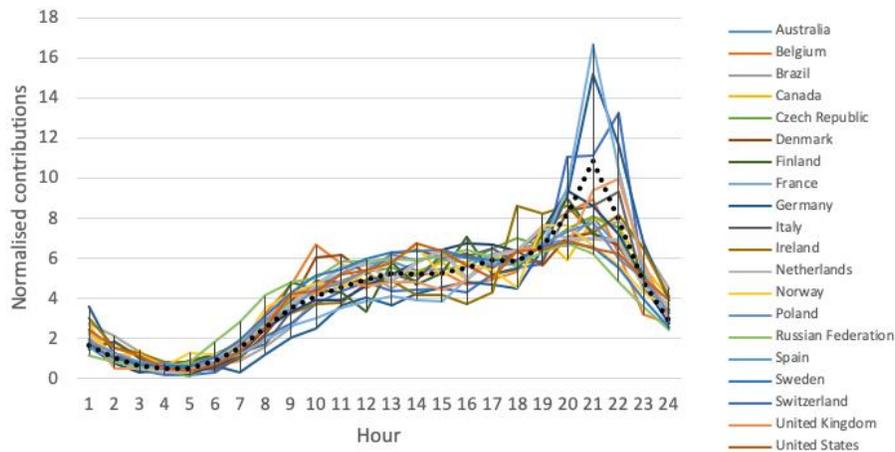

**Figure 7:** Time of first contribution, when volunteers make their first, and for many their only, classification

Gender balance across projects

Overall the gender imbalance among the contributors to the Zooniverse projects is large. Among all the contributions, where the gender of the contributor could be determined with high confidence, only about 30% are committed by female contributors. Whilst the share of female citizen scientists' contributions is considerably larger than similar collaborative projects such a Wikipedia [43] and free/libre/open source software development [44], there is still a large gap of about 40% between the two main genders. Figure 8 shows the percentage of female volunteers foe each country (see Supplementary Information for the full list).

Female volunteers are underrepresented in most countries. In many countries, women make up less than one-third of number of volunteers whose gender is known. The female ratio of participation in the UK and Australia, for instance, is 25 per cent, while the figures for US, Canada and Germany are between 27 and 30 per cent. What is notable here is we see no clear correlation between these ratios and the estimated percentage of female authors in each country [45] with India, Iran, and Ukraine among the countries with the larger share of female contributors in Zooniverse.

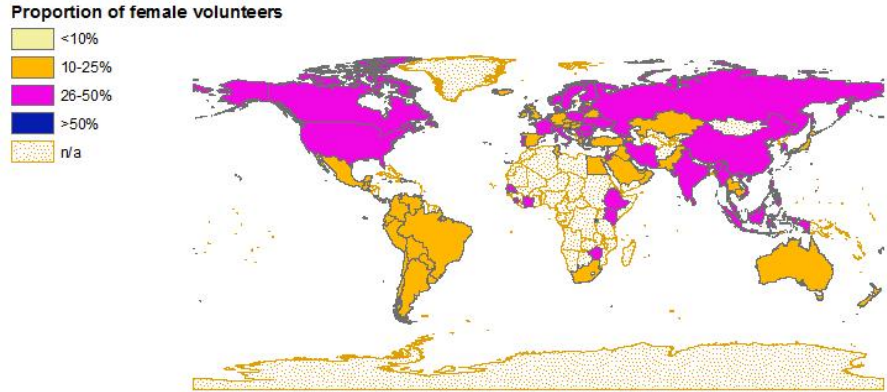

**Figure 8:** ratio of female contributors to Zooniverse projects.

An explanation for this deviation can be that in countries where opportunities for formal participation in knowledge generation activities are sparse for women, there is a larger interest among female citizen scientists to participate through the informal and more accessible environment of citizen science projects.

Another pattern reported for the gender imbalance among formal scientists is the differences between different disciplines with life sciences and social sciences hosting more female scientists compared with physical and mathematical sciences [46]. To test if we see a similar pattern among citizen scientists, we calculated the share of female contributors per project. The top three projects each with more than 50% female contributors are *Snapshot Serengeti*, *Whale FM*, and *Take notes from Nature*, all related to wildlife and nature. On the other hand, the two project with the least number of female scientists (less than 20%) are *Planet Hunters* and *Galaxy Zoo*, both in the field of astronomy. It is notable that whilst the overall pattern of gender imbalance among formal scientists across countries is not preserved among citizen scientists, the uneven distribution of genders among different fields of science are replicated. According to an NSF report, the share of women in engineering is 13%, computer and mathematical sciences 25%, but they are well-represented in the social sciences 58%, and biological and medical sciences 48%.

## Discussion

The growing popularity of citizen science online has a number of implications. Citizen science acts as a bridge between the science community and the public, as scientific institutions have typically kept a distance from the larger society and scientific processes are obscured behind black boxes and ivory walls. It provides volunteers, who normally will not have access to scientific data and research projects, opportunity to become involved in knowledge production. In the process, citizen science helps participants increase their scientific literacy and develop positive attitudes towards science [8]. Indeed, "carefully designed citizen-science projects can be successful environments for increasing adult knowledge of factual science" [5]. Additionally, by leveraging on knowledge creation and sharing models of crowdsourcing [47] and peer-production [48], citizen science bypasses the problems of funding and human resource shortages. It is by harnessing crowd power and engaging in these new forms of scientific collaboration that online citizen science can help facilitate and accelerate scientific discovery.

Our research examined the spatiotemporal patterns of activity in citizen science on the Zooniverse platform. We found that volunteers are unevenly distributed around the world. They are found predominantly in North America and Europe, and in small numbers in the rest of the world. These variations can be explained by socio-economic factors. Over 80 per cent of the variations are due to the combined effects of a country's population, wealth (GDP per capita), Internet connectivity and rate of school enrolment. National emphasis on science, as represented by spending on research and development, the number of researchers employed and publication records, also have an effect on citizen science, accounting for 73% of variation in citizen science activity. By identifying these trends and the factors that produce them, policy makers, scientific institutions as well as citizen science developers can pinpoint strategies to increase involvement in science, one of them through promoting participation in informal science projects such as citizen science.

We also studied the temporal patterns of activity on the Zooniverse, and the results demonstrate that volunteers tend to be most active during the evening, which corresponds with a typical household's leisure time. First time volunteers are also most likely to start classifying during the same time intervals of the day.

Studying the gender imbalance among contributors, we see a higher participation from female citizen scientists compare to Wikipedia and open software development projects, however, still a large gap between the number of male and female contributors is present. We observe smaller gaps in countries with larger gender imbalance in more formal research professions, which suggest citizen science projects can practically play the role of an informal channel for females with strong interest in science where more formal channels are less accessible.

Our findings represent initial forays into research to understand the influence of time and place on involvement in scientific knowledge production. Further studies are needed to discover not just the national-level factors for varying levels of citizen science involvement, but also variables that affect participation on a micro-scale. Doing so would help identify ways to best tap into the vast reserves of interest, time and effort and channel them towards scientific pursuits that the public can contribute to.

## Data Availability
All the dataset used in this research is available at https://doi.org/10.5281/zenodo.583182.

## Authors' Contributions
K.I., S.K., and T.Y. Analyzed the data. T.Y. designed the study and secured the funding. K.I. and T.Y. drafted the manuscript. All authors contributed to writing the manuscript and gave final approval for publication.

## Competing Interest
The authors declare no competing interests.

## Funding
This publication arises from research funded by the John Fell Oxford University Press (OUP) Research Fund, grant no. 132/126. T.Y. was partially supported by the Alan Turing Institute under the EPSRC grant no. EP/N510129/1.

## Acknowledgement
We thank Robert Simpson, Grant Robert MacKinnon Miller, and Chris Lintott from Zooniverse for sharing the data and insightful discussions. We thank Ksenia Musaelyan for comments on the manuscript.